# A single layer hydrogen silsesquioxane (HSQ) based lift-off process for germanium and platinum

V. Passi[a], A. Lecestre[b], C. Krzeminski[b], G. Larrieu[b], E. Dubois[b], and J.-P. Raskin[a]

[a] Université catholique de Louvain (UCL), Electrical Department, Maxwell Building,
Place du Levant, 3, B-1348 Louvain-la-Neuve, Belgium

[b] Institut d'Electronique, de Microélectronique et de Nanotechnologie (IEMN)
Silicon Microelectronics Group, Cité Scientifique, Avenue Poincaré,
BP 60069, F-59652 Villeneuve d'Ascq Cedex, France

**Abstract** - Primarily used as etch mask, single layer hydrogen silsesquioxane has never been investigated for lift-off technique. In this article, we propose a new technique where a single layer of hydrogen silsesquioxane, a negative tone electron beam resist, is used to make lift-off of germanium and platinum. Removal of exposed hydrogen silsesquioxane is tested for various concentrations of hydrofluoric acid. Ultrasonic agitation is also used to reduce the formation of flakes due to accumulation of matter (evaporated metal in our case) along the sidewalls of the lift-off narrow slots. Results demonstrate potential in applying the hydrogen silsesquioxane as a negative tone lift-off resist to pattern nanometer scale features into germanium and platinum layers.

**Keywords**: electron beam lithography, hydrogen silsesquioxane (HSQ), lift-off technique, platinum deposition, germanium deposition, tetramethylammonium hydroxide (TMAH).

## I. Introduction

Lift-off is a relatively simple and easy method for making metal patterns on substrate, especially for metals such as platinum, tantalum, and nickel, which are difficult to etch by conventional methods that involve wet chemical or dry reactive ion etching. So far, the lift-off technique has been applied to positive tone electron beam (e-beam) resists such as poly methyl methacrylate (PMMA) [1]-[4] and ZEP 520-12 [5]. A single layer positive tone e-beam resist process is the simplest to implement and involves only one lithography step. However, the drawbacks of this approach are; the positive tone resists are relatively more expensive than negative tone [6] and the resist removal step that can result in rough sidewall profile of the metal. During deposition, metal film could be deposited on the vertical sidewall of the resist due to insufficient undercut and continue to adhere to the substrate following resist removal. This sidewall may peel-off during subsequent processing steps, resulting in shorts and particulate matter and/or it may collapse and interfere with the ensuing processing steps. Thus, a sufficient undercut would help in avoiding rough sidewall profile of metal. Fig. 1a shows the ideal undercut profile for a lift-off using single layer positive tone resist. There are two methods to obtain an undercut; count on proximity effect, and use a bilayer resist process [7]. The proximity effect determination is a complex procedure depending on numerous parameters such as exposure dose, electron energy, substrate material, resist thickness, development time, development temperature, etc., to name a few. The bilayer process is a relatively easier method that helps in formation of an undercut resulting in a reliable lift-off. A bilayer process consists of two resist layers of different molecular weight spin coated one over the other. During exposure, the bottom layer of low molecular weight being more sensitive to electrons is exposed more than the top layer of higher molecular weight, which is relatively less sensitive. During development, the bottom layer is developed at a faster rate than the top layer resulting in formation of a structure similar to an undercut Fig. 1b. This process however is more complex and of higher cost than a single layer process. In this article, we present a lift-off process using a single layer negative tone e-beam resist, hydrogen silsesquioxane (HSQ) for germanium and platinum.

**II. Lift-off process using positive and negative resists to pattern narrow openings in metal**

In order to fabricate narrow lines of materials, which are difficult to etch by either wet chemical or reactive ion etching, lift-off technique is used. The two main advantages of lift-off process are:
i) It does not require etch selectivity between the film to pattern and the substrate since no etching is required;
ii) Any material or stack of materials can be used as long as they can be evaporated at relatively low temperature (temperature at the wafer surface should be less than the glass transition temperature of the resist to avoid resist overflow).

In order to create narrow openings in metal using lift-off process, a negative tone resist is more beneficial than a positive tone resist in terms of time and cost of exposure. The process steps to obtain narrow openings in metal using positive tone resist are explained briefly. Positive resist layer is spin coated on the substrate. Exposure is done in regions where the metal is to be deposited, followed by development. This results in resist layer remaining in the region where narrow openings in metal are desired. Deposition of metal is then done followed by removal of resist using remover. These steps are illustrated in Fig. 2. As can be seen above this step requires exposure of large areas thereby increasing the time and cost of exposure.

To obtain the same using negative tone resist the process steps are similar to the process steps using positive resist Fig. 2, but the significant advantage being that here the area to be exposed is much less, thereby comparatively reducing the time and cost.

In this article, negative tone resist hydrogen silsesquioxane (HSQ) layer is exposed and developed, followed by deposition of germanium/platinum. Finally, lift-off is performed to obtain narrow openings.

**A. Hydrogen silsesquioxane (HSQ) as a negative tone lift-off resist**

Hydrogen silsesquioxane commonly known as HSQ or Flowable Oxide (FOx), is used as a low-k dielectric for back-end metal interlayer isolation [8]-[9] and has excellent planarization and gap filling capabilities [10]-[11]. We use negative tone electron beam resist HSQ for our application owing to the following advantages that it offers: (i) HSQ can be applied using standard spin coating techniques, (ii) presents a high resolution [12]-[13], (iii) it features good resistance to plasma etching and can be used directly as an etch mask for pattern transfer, (iv) when exposed with e-beam or ultraviolet it approaches the structure of $SiO_x$ [10], [14], (v) its thickness is uniform and well-controlled and (vi) it presents a very low defect density. HSQ-based lift-off process requires the use of hydrofluoric acid (HF) for the removal of the exposed patterns. One disadvantage associated to this is the incompatibility with an underlying layer composed of silicon-dioxide. However, since the etch rate of HSQ in HF is much larger (~ 800 nm/min in HF-1%) than that of e.g. plasma enhanced chemical vapor deposited (PECVD) or thermal $SiO_2$, etching depth of $SiO_2$ can be controlled by varying the etch time and HF-based solution concentration. Another disadvantage is that HF attacks some metals [15]. In our experiments, we do not face that issue since germanium and platinum layers are considered.

**B. Patterning of germanium and platinum**

Germanium, being a group IV semiconductor, shares many similar properties with silicon and is compatible with silicon microfabrication. Germanium is foreseen as an interesting layer for building Micro-Electro-Mechanical Systems (MEMS) in post CMOS process [16], due to its unique complimentary characteristics such as low temperature physical vapour deposition in the amorphous state, excellent etch selectivity to materials that are commonly used in silicon micromachining, such as silicon-dioxide, silicon nitride and aluminium. Being resistant to HF, nanometer scale features can be defined by lift-off using HSQ-based electron beam lithography. Moreover, the excess of germanium can be easily removed using preheated hydrogen peroxide solution (at 50°C) without causing damage to silicon or silicon-dioxide, and the deposition of germanium at low temperature (less than 150°C) does not cause any resist burning.

Germanium can also be used as an implantation mask owing to its stopping power to define buried implanted layers. The change of electrical properties of buried layers by implantation is of interest when reduction of the ohmic contact is desired [17]-[19]. The change of chemical properties of buried layers by implantation is also of interest to create nanometer scale self-aligned gates for a planar double gate MOSFET by tuning the etch rate of implanted oxide in vapour phase hydrofluoric acid [20]-[23]. It is worth noting that metal like platinum can be patterned as demonstrated hereafter by a lift-off process of HSQ. This last remark is of special interest as platinum is a widely used metal layer for building micro and nanometer scale metallic lines or nanometer spaced interdigitated electrodes in the field of biosensors [24]-[27], for instance.

### III. Experiments

Figure 3 shows the process flow of lift-off using HSQ for germanium/platinum. The starting substrate is a p-type bulk silicon wafer. The process begins with a standard cleaning step followed by HF-1% dip to remove chemical oxide followed by de-ionized (DI) water rinse. Dry oxidation is performed at 1100°C to obtain a 200 nm-thick silicon-dioxide layer. Flowable Oxide (FOx-16) is subsequently spin-coated at a speed of 1000 rpm for 60 s, baked at 90°C for 4 minutes resulting in a thickness of 500 nm (as measured by ellipsometry). Electron beam exposure is performed using a VISTEC EBPG 5000+ beam writer at 50 keV to obtain 0.1, 0.3, 0.5, 0.7, and 1 µm-wide isolated lines. Resist is developed in tetramethyl ammonium hydroxide (TMAH) – 25% solution for 1 minute followed by DI water rinse for 2 minutes. Scanning electron microscope imaging is then performed to determine the width of the patterned structures. A 300 nm-thick germanium is deposited by evaporation. For platinum process, 50 nm-thick platinum is deposited by evaporation. Lift-off is done by dipping the sample in HF solution. Various concentrations of HF solutions and various etching times are tested to optimize the etching of HSQ in order to obtain good quality lift-off results.

**IV. Results and discussion**

Cross section of HSQ lines of varying widths after exposure and development are shown in Fig. 4-a through 4-d. Dose determination is done by varying exposure dose, from 1500 µC/cm² to 3000 µC/cm² in steps of 100 µC/cm², to obtain lines widths from 100 nm to 1 µm. Line widths are measured using Scanning Electron Microscopy (SEM) image for each dose value. Finally, the dose value of 2500 µC/cm² has been selected since it gives the desired dimension for each width. Fig. 5-a and 5-b show the under etching of silicon-dioxide resulting from a lift-off in concentrated HF (50%) for 3 minutes and 30 s, respectively. Alternatively, Fig. 5-c and 5-d show the cross section after lift-off in a diluted HF solution (1%) for time of 15 and 10 minutes, respectively. The underetch observed in Fig. 5-c and 5-d is further reduced by reducing the lift-off time to 2 minutes. In addition, the flakes on the sidewalls are removed with the help of ultrasonic agitation. Fig. 5-e and 5-f exemplify this optimized process for lift-off of HSQ. The presence of the slope in the sidewalls accounts to the shadowing effect from germanium deposited on top surface of HSQ. One way of reducing this effect is to decrease the thickness of the deposited germanium layer.

Fig. 6-a through 6-d show the cross section SEM image of openings with varying width after lift-off of HSQ for germanium.

For a platinum based process, top-view of HSQ lines after exposure and development for various widths are shown in Fig. 7-a through 7-d. The spacing between the lines is varied from 300 nm to 2 µm. Fig. 8-a through 8-d show narrow openings in platinum from 100 nm down to 25 nm after HSQ lift-off. The use of ultrasonic agitation results in better sidewall definition by reducing the metal flakes as is also observed for germanium. By tuning the HF concentration and the duration of lift-off, the unintentional etching of the silicon-dioxide layer lying on top of the silicon substrate is controlled.

## V. Conclusion

We have demonstrated a new process to perform a lift-off using single layer negative tone resist, hydrogen silsesquioxane (HSQ), for germanium and platinum.

The required HSQ thickness of 500 nm is obtained with the spin coat parameters; speed 1000 rpm for 60 s, bake temperature 90°C for 4 minutes. HSQ widths of 0.1, 0.3, 0.5, 0.7, and 1 µm are achieved with an exposure dose of 2500 µC/cm². Different concentrations and etching duration of hydrofluoric acid (HF) to remove the exposed HSQ are tested. Influence of the ultrasonic agitation on etched profile geometry and debris presence is evaluated. Best lift-off results are obtained when 1% of HF for 2 minutes with presence of ultrasound agitation is used. Narrow openings of 100 nm for germanium lift-off case and 25 nm for the platinum lift-off case are defined by e-beam lithography without any proximity correction.

**Figure Captions**

**Figure 1**: Undercut profile for lift-off using (a) single layer and (b) bilayer positive tone resist.

**Figure 2**: Comparison of positive and negative tone process for obtaining narrow openings in metal.

**Figure 3**: Process flow of HSQ-based lift-off for germanium/platinum.

**Figure 4**: Cross section SEM images of HSQ lines with a width of (a) 100, (b) 300, (c) 500, and (d) 700 nm after e-beam exposure and development in TMAH-25% for 1 min followed by DI rinse. These patterns are exposed with 50 keV energy at a dose of 2500 µC/cm² without proximity correction.

**Figure 5**: Cross section SEM images showing lift-off of 300 nm-thick Ge layer in various wet HF-based solutions: (a) HF 50% for 3 min, (b) HF 50% for 30 s, (c) HF 1% for 15 min, (d) HF 1% for 10 min, (e) HF 1% for 2 min, together with the use of ultrasonic agitation to reduce the sidewall flakes of germanium for an opening of 1 µm and (f) HF 1% for 2 min, with ultrasonic agitation for an opening of 100 nm.

**Figure 6**: Cross section SEM images after lift-off of HSQ with 300 nm-thick germanium using HF – 1% for 2 minutes with ultrasonic agitation. Openings of (a) 100, (b) 300, (c) 500, and (d) 700 nm-width are patterned to access to a silicon-dioxide layer of 200 nm-thick.

**Figure 7**: Top view SEM images of HSQ narrow lines of various widths and spaced from 300 nm to 2 µm, for platinum based process, after exposure and development: (a) 25 nm, (b) 50 nm, (c) 75 nm and (d) 100 nm wide HSQ lines.

**Figure 8**: Top view SEM images of openings in platinum for different spacing: (a) Openings of 25 nm width with spacings of 500 nm and 1 µm, (b) Openings of 50 nm width and spacings from 300 nm to 1 µm, (c) Openings of 75 nm width with spacings from 300 nm to 2 µm and (d) Openings of 100 nm width with spacings from 300 nm to 1 µm.

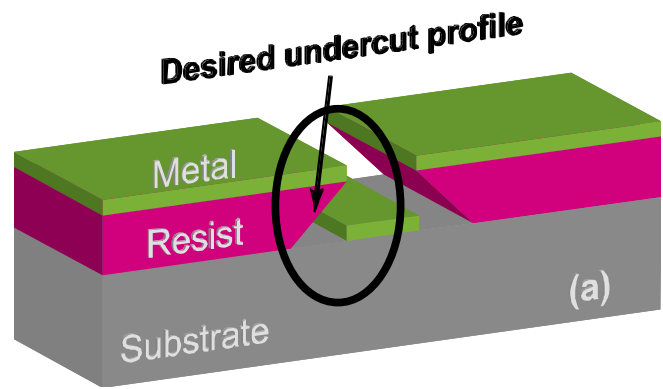

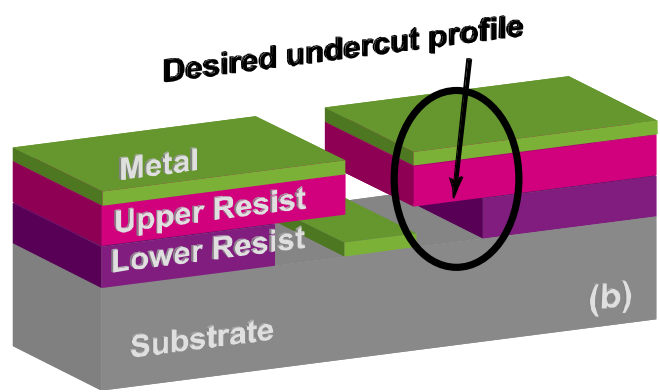

**Figure 1**: Undercut profile for lift-off using (a) single layer and (b) bilayer positive tone resist.

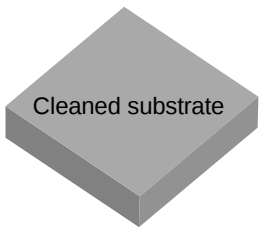
Cleaned substrate

Resist spin coating

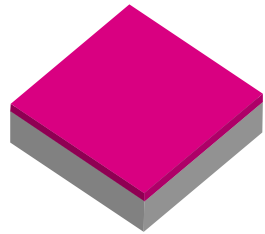

**Negative tone resist process**

**Positive tone resist process**

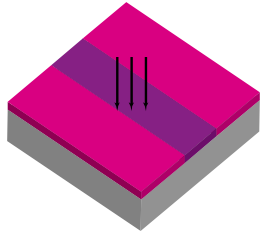
Exposure
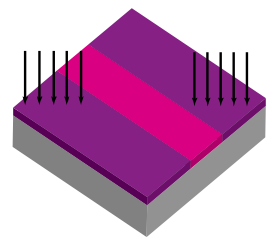

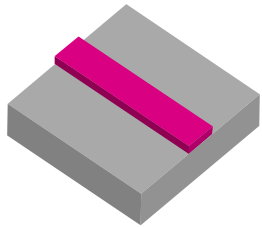
Development
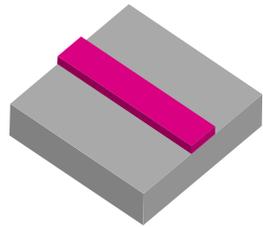

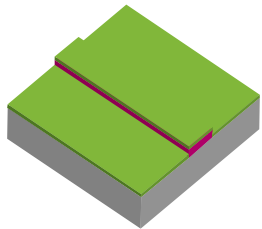
Metal evaporation
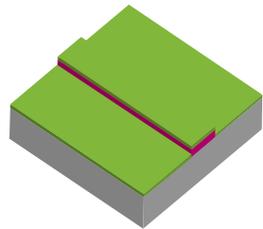

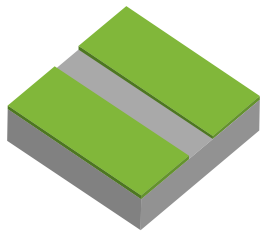
Metal Lift-off
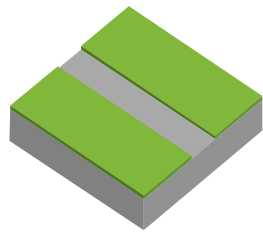

Metal
Resist
Substrate

**Figure 2**: Process flow to get narrow opening in metal using positive and negative tone resists.

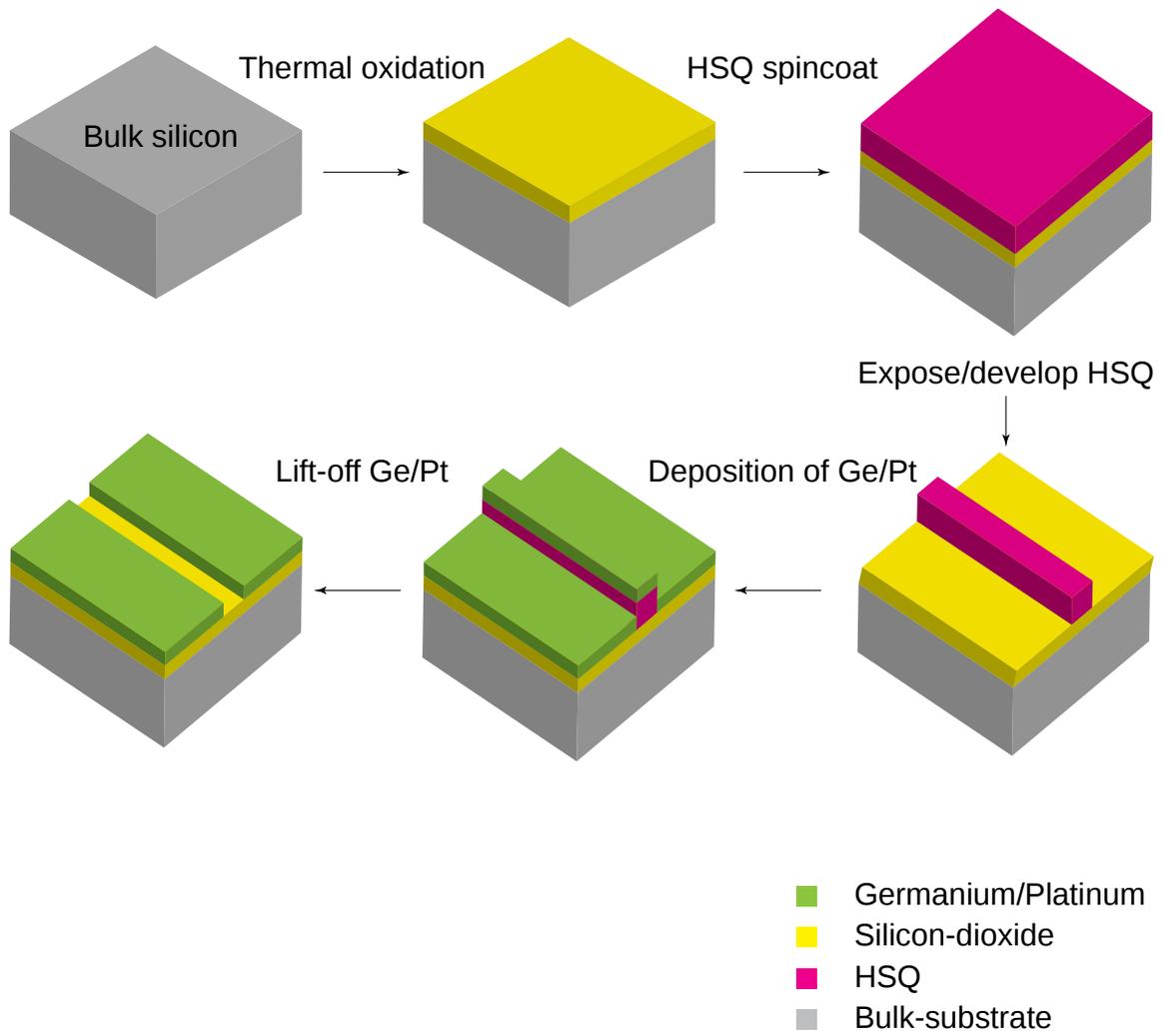

**Figure 3**: Process flow of HSQ based lift-off for germanium/platinum.

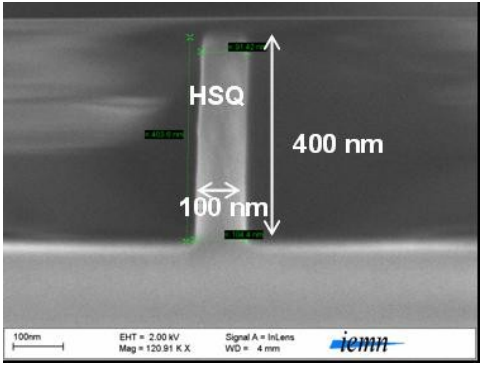
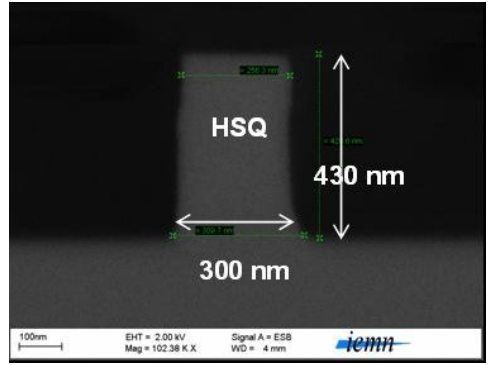

(a)             (b)

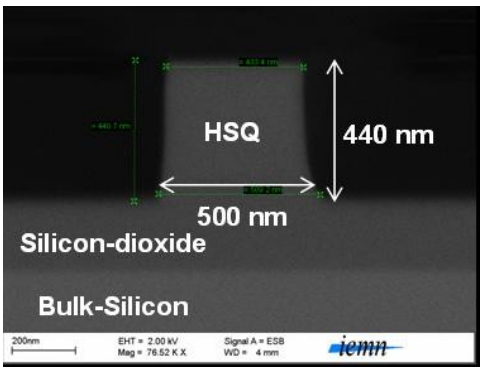
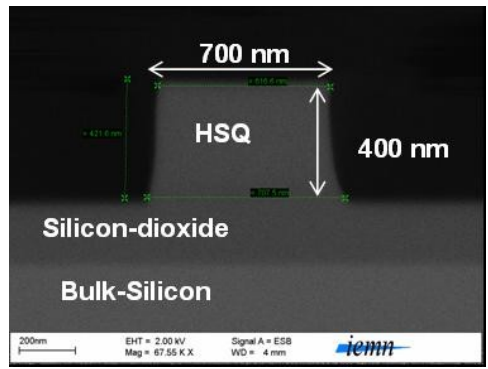

(c)             (d)

**Figure 4**: Cross section SEM images of HSQ lines with a width of (a) 100, (b) 300, (c) 500, and (d) 700 nm after e-beam exposure and development in TMAH-25% for 1 min followed by DI rinse. These patterns are exposed with 50 keV energy at a dose of 2500 µC/cm² without proximity correction.

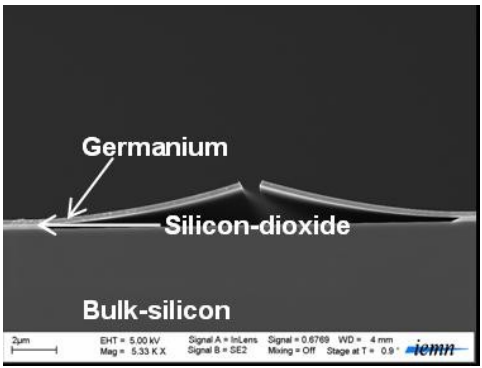 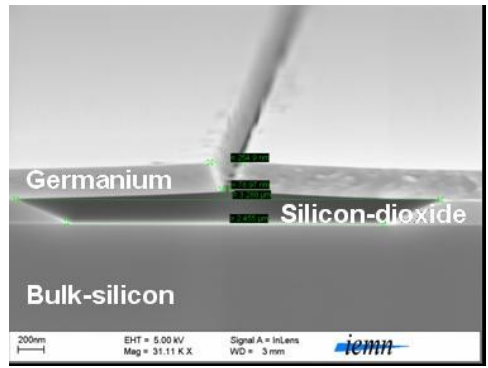

(a) (b)

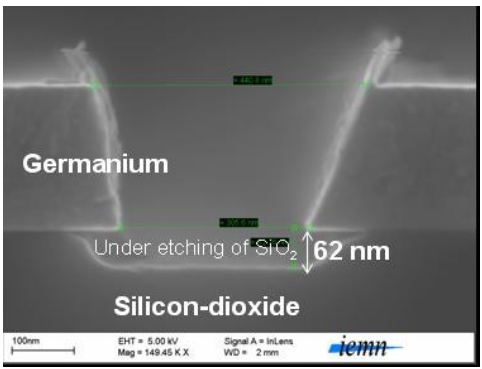 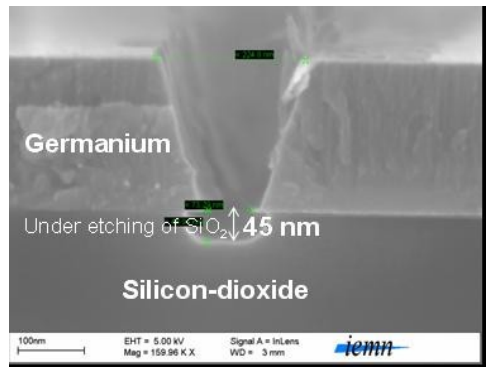

(c) (d)

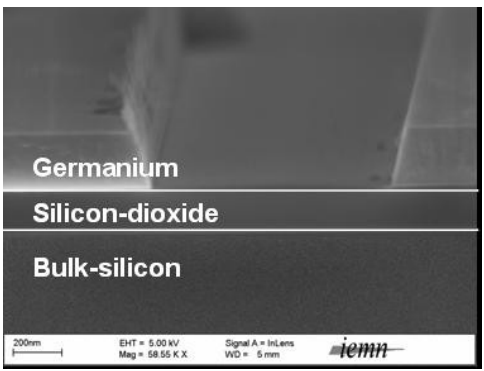 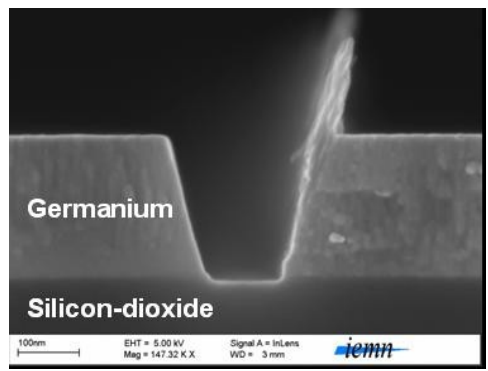

(e) (f)

**Figure 5**: Cross section SEM images showing lift-off of 300 nm-thick Ge layer in various wet HF-based solutions: (a) HF 50% for 3 min, (b) HF 50% for 30 s, (c) HF 1% for 15 min, (d) HF 1% for 10 min, (e) HF 1% for 2 min, together with the use of ultrasonic agitation to reduce the sidewall flakes of germanium for an opening of 1 µm and (f) HF 1% for 2 min, with ultrasonic agitation for an opening of 100 nm.

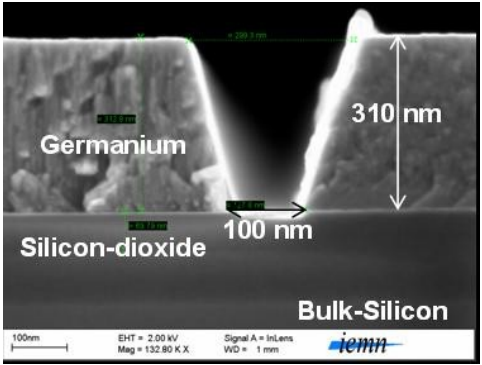 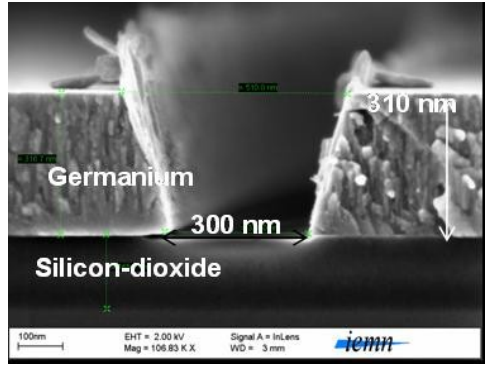

(a)　　　　　　　　　　　　　　　　　　(b)

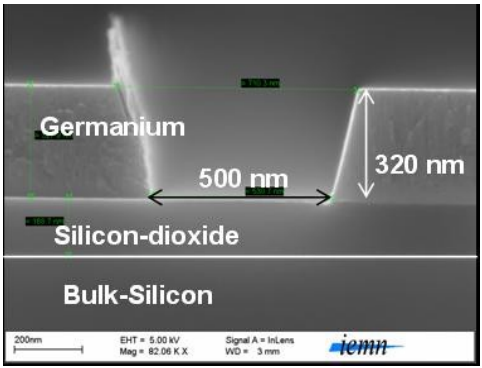 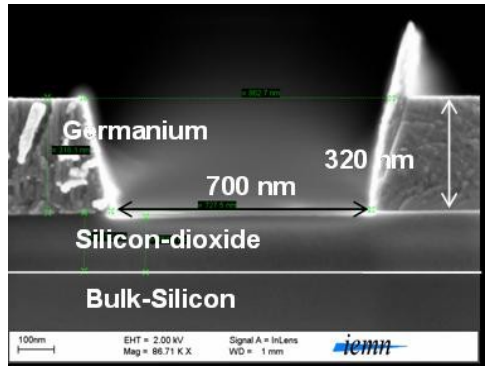

(c)　　　　　　　　　　　　　　　　　　(d)

**Figure 6**: Cross section SEM images after lift-off of HSQ with 300 nm-thick germanium using HF – 1% for 2 minutes with ultrasonic agitation. Openings of (a) 100, (b) 300, (c) 500, and (d) 700 nm-width are patterned to access to a silicon-dioxide layer of 200 nm-thick.

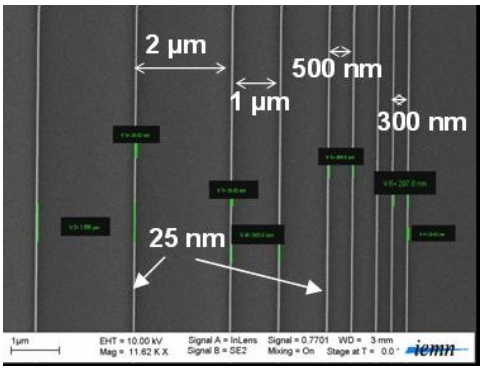 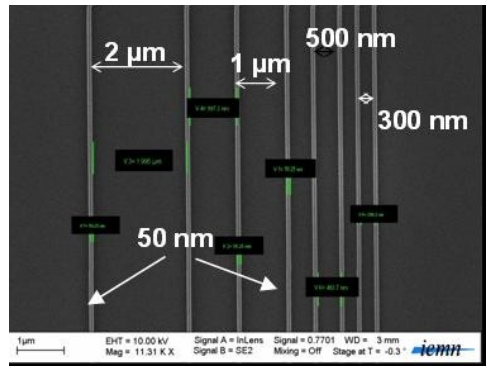

(a) (b)

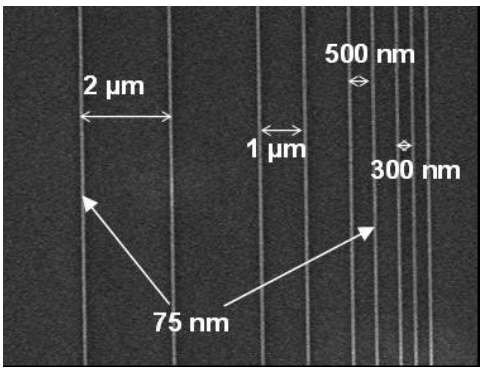 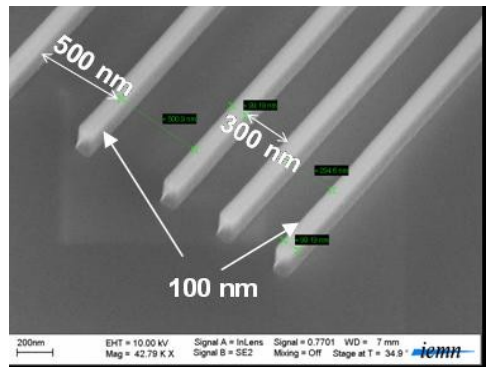

(c) (d)

**Figure 7**: Top view SEM images of HSQ narrow lines of various widths and spaced from 300 nm to 2 µm, for platinum based process, after exposure and development: (a) 25 nm, (b) 50 nm, (c) 75 nm and (d) 100 nm wide HSQ lines.

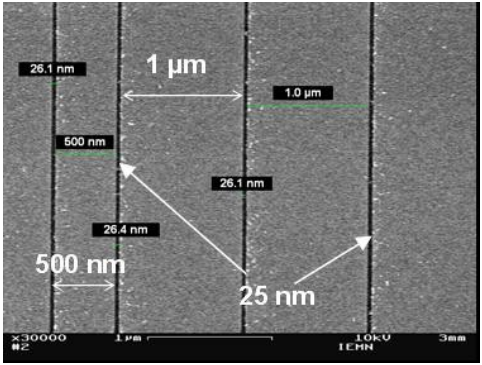

(a)

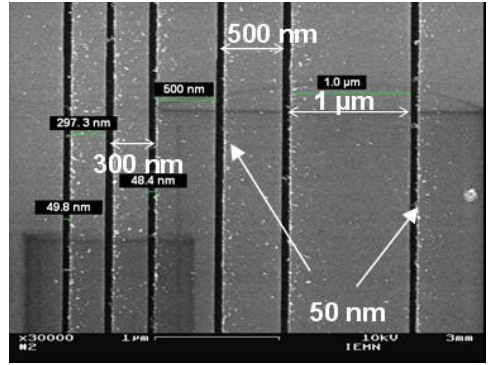

(b)

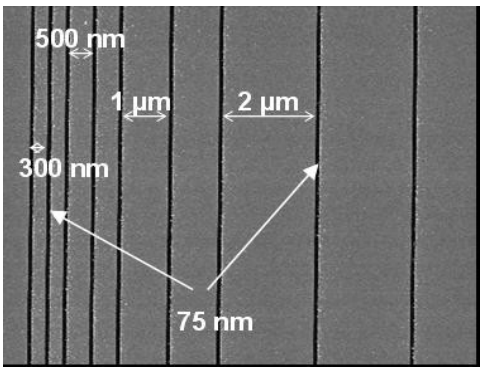

(c)

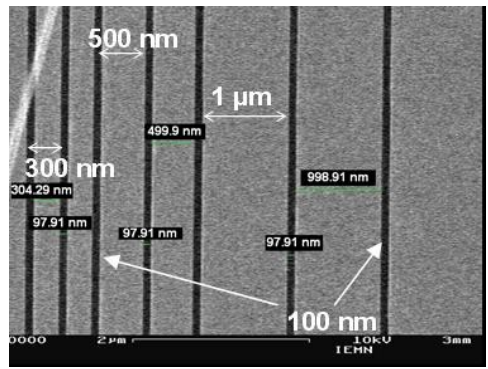

(d)

**Figure 8**: Top view SEM images of openings in platinum for different spacing: (a) Openings of 25 nm width with spacings of 500 nm and 1 µm, (b) Openings of 50 nm width and spacings from 300 nm to 1 µm, (c) Openings of 75 nm width with spacings from 300 nm to 2 µm and (d) Openings of 100 nm width with spacings from 300 nm to 1 µm.